\documentclass{IEEEtran}
\usepackage{generic}
\usepackage{algorithmic}
\usepackage{graphicx}
\usepackage{textcomp}

\usepackage{threeparttable}
\usepackage{makecell}

\usepackage{amsmath,amssymb,amsfonts}
\usepackage{physics}
\usepackage{siunitx}
\usepackage{adjustbox}

\usepackage{cite}
\usepackage[hidelinks]{hyperref}

\def\BibTeX{{\rm B\kern-.05em{\sc i\kern-.025em b}\kern-.08em
    T\kern-.1667em\lower.7ex\hbox{E}\kern-.125emX}}
\begin{document}
\title{Three mechanistically different variability and noise sources in the trial-to-trial fluctuations of responses to brain stimulation}

\author{Ke Ma, 
        Siwei Liu, 
        Mengjie Qin,
        and Stephan M.\ Goetz
}
\maketitle

\begin{abstract}
Motor-evoked potentials (MEPs) are among the few directly observable responses to external brain stimulation and serve a variety of applications, often in the form of input--output (IO) curves. Previous statistical models with two variability sources inherently consider the small MEPs at the low-side plateau as part of the neural recruitment properties. However, recent studies demonstrated that small MEP responses under resting conditions are contaminated and over-shadowed by background noise of mostly technical quality, e.g., caused by the amplifier, and suggested that the neural recruitment curve should continue below this noise level. This work intends to separate physiological variability from background noise and improve the description of recruitment behaviour. We developed a triple-variability-source model around a logarithmic logistic function without a lower plateau and incorporated an additional source for background noise. Compared to  models with two or fewer variability sources, our approach better described IO characteristics, evidenced by lower Bayesian Information Criterion scores across all subjects and pulse shapes. The model independently extracted hidden variability information across the stimulated neural system and isolated it from background noise, which led to an accurate estimation of the IO curve parameters. This new model offers a robust tool to analyse brain stimulation IO curves in clinical and experimental neuroscience and reduces the risk of spurious results from inappropriate statistical methods. The presented model together with the corresponding calibration  method provides a more accurate representation of MEP responses and variability sources, advances our understanding of cortical excitability, and may improve the assessment of neuromodulation effects.
\end{abstract}

\section{Introduction}
Motor-evoked potentials (MEPs) are among the few directly measurable and observable acute responses to transcranial magnetic stimulation (TMS) of the brain \cite{rossini2015_noninvasive_stimulation, goetz2017development}. They therefore serve as a significant biomarker for many applications, such as probing excitability changes through neuromodulation or through the motor threshold as a stimulation reference for almost all other stimulation procedures \cite{rossini2015_noninvasive_stimulation,ma2023_MT_distribution, MCCONNELL2001454, OUPneuronavigation}. Multiple MEPs administered over a range of stimulation strengths typically form an s-shaped curve with a low-side and a high-side plateau, also named the motor input--output (IO) curve or recruitment curve, is a well-known and important detection measurement for assessing changes in the motor system \cite{ridding1997_MEP_cortical_excitability, peterchev2013_pulse_width_IO, goldsworthy2016_TBS_IO}.
Peripheral neural and neuromuscular stimulation forms similar IO curves \cite{goetz2022prediction,goetz2011comparison,bickel2011motor,schearer2012optimal}.
IO curves are commonly fitted using ordinary least-square regression of sigmoidal curves, such as Boltzmann- and Hill-type functions, which intrinsically assume additive Gaussian errors on top of the MEP size \cite{moller2009_hysteresis_IO, devanne1997_IO_curve_model,alavi2019optimal,10061537}.

\par
However, this simple method is insufficient to model the inherently variable and noisy MEP features. The MEP in response to the brain and certain types of spinal stimulation can exhibit high levels of intra-individual trial-to-trial variability and rather intricate, highly skewed, heteroscedastic distributions \cite{van1996_MEP_cortical_excitability, kiers1993_MEP_variability, capaday2021_variability_muscle}. Therefore, the least-square fitting of a linear sigmoid would lead to asymmetric residuals and is prone to spurious results, which could cause systematic errors in the IO curve parameters and derived biomarkers. Instead, logarithmic normalisation of the MEP distribution was strongly recommended and has been suggested to improve stability as well as reduce model bias when calibration \cite{nielsen1996logarithmic, wassermann2002_MEP_variation}.

\par
Beyond normalisation, available statistical models could identify and isolate several widely independent sources of variability involved in IO curves. One additive source on the stimulation side appears to primarily affect the curve in the \textit{x} direction, i.e., the stimulation-strength direction, and effectively modulates the neuronally effective stimulus strength. Another variability source is rather independent of the stimulation strength but rather depends on the response size as it multiplicatively changes the MEPs in the \textit{y} direction, i.e., the MEP-amplitude direction \cite{goetz2012model_MEP_variability}. This multiplicative variability decreases with the MEP size and would logically be zero without any neural activity.

The model with those two variability sources could quantitatively describe the size of both variability sources. For example, cortical excitability fluctuations would be exclusively represented by the additive variability source and the spinal and muscular pathways by the multiplicative source \cite{jung2010_navigated_MEP, kiers1993_MEP_variability, goetz2014_statistical_MEP_model}. In comparison with the conventional least-square regression model, later research refined the model and confirmed its statistically better description of the observed IO characteristics of changing distribution spread and skewness with stimulus strength \cite{goetz2014_statistical_MEP_model}.

\par
Apart from fluctuations in the corticospinal system,  also background noise that is entirely independent of the MEP significantly affects the shape of the motor IO curves. This background noise  represents additional fluctuations that also emerge in a recording when the electrodes are on the skin but no MEP is elicited. It most likely comes from the amplification system, known as thermal noise, and the interface between the skin and the electrode as well as the impedance of the path through the tissue \cite{Letzter1970_Amplifier, dunnewold1998_electrode_noise}. Additionally, the bioelectric activity from other muscle units that are not targeted can also contribute to the noise floor \cite{de2010_EMG_noise}.  For low stimulus strengths, the MEP responses seem to disappear, and this measurement noise dominates the lower plateau of the IO curve \cite{goetz2018_noise_floor_detection}.

A more sensitive matched-filter signal detector instead of a conventional peak-to-peak reading can decrease the effective noise level and therefore also the lower plateau of the MEP IO curve by as much as an order of magnitude \cite{goetz2018_noise_floor_detection, li2022_noise_floor_detection}. Thus, technical rather than physiological aspects determine the level of the low-side plateau. Since it is primarily determined by technical factors, it appears reasonable to not include it in the IO curve or consider it even as a biomarker. From another perspective, these studies suggested that, due to the limitations of the amplifier and detection techniques, the background noise conceals weak MEP responses also in cases when the neural system actually does respond to the small stimuli. Hence, although capable of fitting and explaining the stimulus-strength-dependent MEP amplitude distributions in previous studies, it appears inappropriate to use an s-shape curve with a lower boundary for simplicity as it lumps both the background noise floor and weak MEP responses together. The assumed size of such a lumped variability would then depend on sampling, i.e., how many samples that are dominated by neural variability and how many dominated by the different background noise are in the dataset. A separation of both should also demonstrate their quantitative deviation in additional to their physical and neurophysiological differences.

\par
Therefore, we propose a new statistical model that separates MEP responses and background noise. The model incorporates three variability sources beyond the two previously identified ones: one sits at the input side and modulates the stimulation site or at locations where the recruitment information is still available due to interactions of various units; one may act at the corticospinal and muscular pathways (multiplicative); and one acts at the measurement site and the equipment (additive). Consequently, this model is called a triple-variability-source model. This separation allows us to independently represent the neural properties without the disturbance from the background noise. Furthermore, the variability sources have rather different modes of action (multiplicative vs.\ additive) and with different distributions. Finally, we decided to remove the sigmoidal recruitment model, specifically its lower plateau, because the lower plateau was found to lack a physiological basis. Instead, it appeared to be an artefact caused by technical issues such as equipment noise and varying electrode impedance during a session, which form this (additive) noise floor below which one could not detect responses. The model can analytically extract hidden information of variability sources over the entire stimulated neural system from the MEP responses. It quantitatively describes the noise floor and identifies the expected MEP amplitudes below the measurement noise floor at lower stimulus strengths.

\section{Methodology}
\subsection{Mathematical framework}
\begin{figure}[h]
    \centering
    \includegraphics[width = \linewidth]{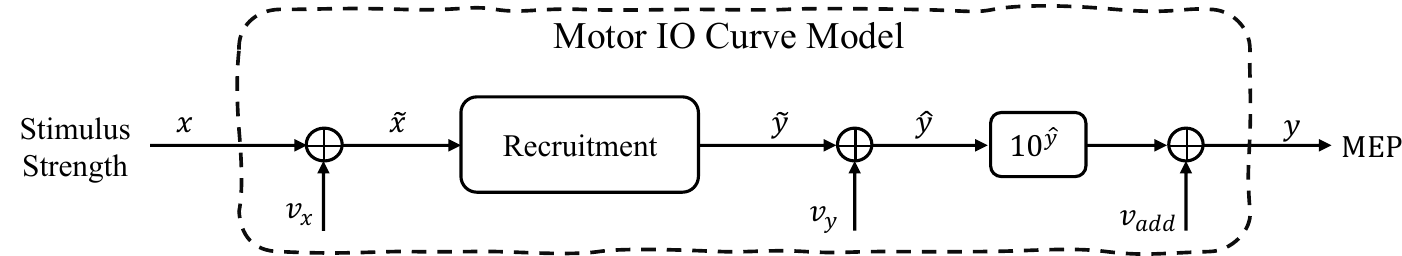}
    \caption{Block diagram of the motor input--output (IO) response model with triple variability sources. Variable $x$ is the stimulus strength, recruitment is the neural recruitment characteristics, $10^{\Tilde{y}}$ is an exponential transformation block, and $y$ is the measured peak-to-peak electromyographic (EMG) amplitude. Independent stochastic variables $v_\textrm{x}$, $v_\textrm{y}$, and $v_\textrm{add}$ with their density functions $g_\textrm{x}$, $g_\textrm{y}$, and $g_\textrm{add}$ represent three different variability sources respectively. $\Tilde{x}$ and $\Tilde{y}$ are respectively the input and output of the characteristic function. MEP: motor-evoked potential.}
    \label{fig: model structure}
\end{figure}

Figure \ref{fig: model structure} depicts the overall structure of this triple-variability-source MEP IO model of the stimulated motor system. The input of the model, $x$, is stimulus strength and its output is MEP amplitude, $y$. The model has two distinct functional blocks. The recruitment block represents how the neural system responds to external stimulation. We used a (logarithmic) logistic function
\begin{equation}
    \Tilde{y} = S(\Tilde{x}) = \log_\textrm{10} \left(\frac{a}{1 + \exp(-b\cdot(\Tilde{x} - c))} \right)\!{},
    \label{equ: recruitment curve}
\end{equation}
where $\Tilde{y}$ is the output of the recruitment function, which will be potentiated later on to practically compensate the logarithm after the effect of the second variability source (see Figure \ref{fig: model structure}), $\Tilde{x} = x + v_\textrm{x}$ is the effective excitation at the input of the recruitment, and all parameters must be positive values. It is worth noting that the recruitment function has no lower boundary. The exponential transformation block translates the neural responses from the logarithmic domain to the normal scale domain to symmetrise the MEP distribution \cite{nielsen1996logarithmic, ma2023_MT_distribution}. The base of the exponential function can be chosen freely because of the logarithmic base change rule. Hence, we used $10$ as the base here for simplicity.

The motor IO curve model has three independent variability sources which act at various locations over the entire neural pathway from the cortex to the muscular motor units. Before the recruitment, the $v_\textrm{x}$ variability source, as an additive noise, acts along the \textit{x} axis of the IO curve and can represent the short-term excitability fluctuations of the neurons directly activated by the stimulus. The output-side variability source $v_\textrm{y}$ before the exponential transformation affects the MEP responses in the logarithmic domain and accordingly has a multiplicative character as shown in Figure \ref{fig: model structure}. This variability source may result from short-term fluctuations in the spinal pathways, the synaptic connection between motor neurons, and the neuromuscular junction and the myocytes \cite{capaday2021_variability_muscle, faisal2008_noise_in_neuron}. Moreover, recent studies demonstrated that the motor system  can still be activated by weak stimulation and suggested that a minimum response for the cortical neuronal target likely exists but seems far less than the background noise floor \cite{goetz2014_statistical_MEP_model, goetz2018_noise_floor_detection, li2022_noise_floor_detection,lazzaro1998magnetic}. Thus, we introduced an additional variability source along the \textit{y} axis, $v_\textrm{add}$, after the exponentiation to represent the additive noise. Therefore, the entire IO curve model becomes
\begin{equation}
\adjustbox{max width=\columnwidth}{$
\begin{aligned}
    y(x) &= v_\textrm{add} + \exp\Biggl( \ln(10) \cdot \biggl( \log_\textrm{10} \Bigl(\frac{a}{1 + \exp\big(-b \cdot (x + v_\textrm{x} - c)\big)}\Bigr) \\
    &\quad + v_\textrm{y} \biggr) \Biggr) \\
    &= v_\textrm{add} + 10^{v_\textrm{y}} \cdot \frac{a}{1 + \exp\big(-b \cdot (x + v_\textrm{x} - c)\big)}. 
\end{aligned}
$}
\label{equ: model}
\end{equation}
This IO curve models the neural recruitment characteristics with a low-side plateau not as part of recruitment but resulting from the background noise, a monotonically growing section representing the increasing population of activated neurons as the stimulus strength increases, and an upper saturation plateau corresponding to the maximum achievable population responses to the sufficiently strong stimulus. Thus, in comparison with previous recruitment curve models, our curve Equation \ref{equ: recruitment curve} no longer uses a lower plateau to represent the background noise floor, for which we used another random variable, $v_\textrm{add}$, instead. Hence, this functional property can independently consider the noise floor and help us easily distinguish it from the small MEP responses in a semi-logarithmic plot.

\subsection{Model parameters}
The triple-variability-source model has two different groups of parameters, one set comes from the  logistic function and the other set from those three aforementioned variability sources. Fitting this model only requires the measured stimulus--response pairs $(x_i, y_i)$. Since we need to estimate the parameters of specific probability density distributions, we applied the maximum likelihood estimation to fit the model using such collected data. In the perspective of conditional probability, Equation \ref{equ: model} explains that, for a given $x$ and a parameter set of the model $\boldsymbol\theta$, the system may output the corresponding $y$ with a certain conditional probability, $f_{y|x}(y | x, {\boldsymbol\theta})$. Calculating the corresponding conditional probability requires consideration of the entire IO model. For a fixed stimulus $x$, the input of the recruitment $\Tilde{x}$ has a probabilistic density function $f_{\Tilde{x}}(\Tilde{x})$, which is equivalent to the density distribution of the variability source $v_\textrm{x}$, $g_\textrm{x}(\Tilde{x} - x)$. The conditional probability density distribution of the output dependent on $x$, $f_{\Tilde{y} | x}(S(\Tilde{x}) | x)$, thus follows
\begin{equation}
     f_{\Tilde{y} | x}(S(\Tilde{x}) | x) \cdot \dd{S(\Tilde{x})} = g_\textrm{x}(\Tilde{x} - x) \cdot \dd{\Tilde{x}},
\end{equation}
where $\dd\cdot$ denotes a traditional differential.

The second variability source $v_\textrm{y}$ involved in $\hat{y}$ (see Figure \ref{fig: model structure}) interplays with $\Tilde{y}$. According to the probability convolution for the addition of two independent probability distributions, the density function of $\hat{y}$ given a fixed $x$ is
\begin{equation}
    \begin{split}
          f_{\hat{y} | x}(\hat{y} | x) & = \int_{\mathbb{R}} g_\textrm{y}(v_\textrm{y}) \cdot     f_{\Tilde{y} | x}(\hat{y} - v_\textrm{y} | x) \cdot \dd{v_\textrm{y}} \\
          & = \int_{\mathbb{R}} g_\textrm{y}(v_\textrm{y}) \cdot g_\textrm{x}(S^{-1}(\hat{y}-v_\textrm{y}) - x) \cdot \\
          &\quad \abs{\dv{\Tilde{x}}{S(\Tilde{x})}\biggr|_{\Tilde{x} = S^{-1}(\hat{y} - v_\textrm{y})}}    \cdot \dd{v_\textrm{y}},
    \end{split}
\end{equation}
where $S^{-1}(\cdot)$ is the inverse function of $S(\cdot)$ and the value of $\dv*{\Tilde{x}}{S(\Tilde{x})}$ is evaluated at point $\Tilde{x} = S^{-1}(\hat{y} - v_\textrm{x})$.

\par
Likewise, the probability density distribution of $10^{\hat{y}}$ after exponentiation follows the conversion
\begin{equation}
    f_{10^{\hat{y}} | x}(10^{\hat{y}} | x) \cdot \dd{10^{\hat{y}}} = 
    f_{\hat{y} | x}(\hat{y} | x) \cdot \dd{\hat{y}}.
\end{equation}
Therefore, the probability density distribution of the output $y$ is
\begin{equation}
    \begin{split}
         f_{y|x}(y|x) &= \int_\mathbb{R} g_\textrm{add}(v_\textrm{add}) \cdot f_{10^{\hat{y}}|x}(y - v_\textrm{add} | x) \cdot \dd{v_\textrm{add}} \\
         &= \int_\mathbb{R} g_\textrm{add}(v_\textrm{add}) \cdot f_{\hat{y}|x}(\log_{10}(y - v_\textrm{add}) | x) \cdot \\
         &\quad \abs{\dv{\hat{y}}{10^{\hat{y}}}\biggr|_{\hat{y} = \log_{10}(y - v_\textrm{add})}} \cdot \dd{v_\textrm{add}}.
    \end{split}
    \label{equ: conditional prob equation}
\end{equation}

Both variability sources $v_\textrm{x}$ and $v_\textrm{y}$ are implemented as independent stochastic elements with Gaussian distribution and respectively obey $\mathcal{N}(0, \sigma_\textrm{x})$ and $\mathcal{N}(0, \sigma_\textrm{y})$, where $\sigma_\textrm{x}$ and $\sigma_\textrm{y}$ are the standard deviations. Previous work suggested that the background noise variability source $v_\textrm{add}$ transformed by a peak-to-peak extraction algorithm is non-Gaussian but obeys a generalised extreme-value distribution, $GEV(k, \sigma_\textrm{add}, \mu_\textrm{add})$, instead \cite{goetz2019_simulation_IO_model}. $k$ controls the distribution shape, $\sigma_\textrm{add}$ the distribution spread, and $\mu_\textrm{add}$ the distribution location \cite{kotz2000_GEV}. Accordingly, the entire model parameter vector, which fully describes a subject's recruitment behaviour to a specific pulse type, is $\boldsymbol{\theta} = [a, b, c, \sigma_\textrm{x}, \sigma_\textrm{y}, k, \sigma_\textrm{add}, \mu_\textrm{add}]$.

\subsection{Data acquisition}
We obtained the experimental data from the literature with nineteen healthy subjects (age $18-45$, $12$ female and $7$ male, all right-handed) \cite{li2022_noise_floor_detection,goetz2018_noise_floor_detection}. The data include single-pulse TMS with various pulse shapes, such as monophasic, reverse monophasic, and biphasic. MEPs were recorded and sampled synchronously to the TMS pulse trigger through surface Ag/AgCl electrodes and an MEP amplifier (K800 with SX230FW pre-amplifier, Biometrics Ltd, Gwent, UK) at $5\,\unit{\kilo\hertz}$ and $16$ bit. The recording starts $200\,\unit{\milli\second}$ before the delivery of the testing TMS pulse, and the whole duration spans $410\,\unit{\milli\second}$.

\begin{figure}[t]
    \centering
    \includegraphics[width=\linewidth]{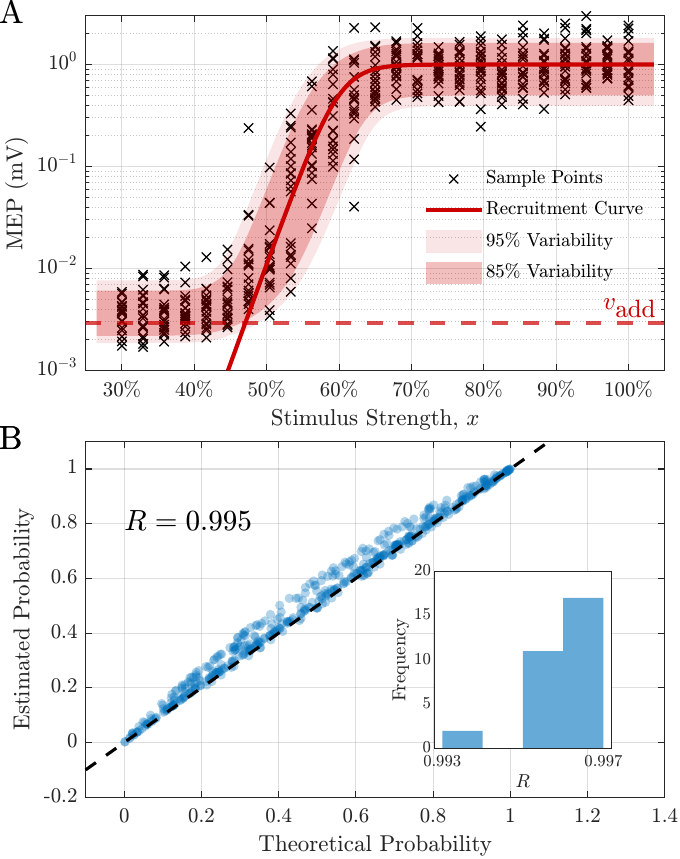}
    \caption{Representative triple-variability-source model calibrated to simulated data with known properties. (\textbf{A}) Calibrated curve model consisting of a mode value of the background noise, $v_\textrm{add}$ (red dashed line) and an expected neuronal recruitment curve (red solid line). The two variability ranges are $95\%$ (shaded with light red) and $85\%$ (shaded with moderate red). (\textbf{B}) Comparison of the cumulative probability between the theoretical models (theoretical probability) and the representative estimated model (estimated probability) with a correlation coefficient of $R = 0.995$. The inset represents the distribution of correlation coefficients between the theoretical and estimated probability for $K = 30$ cross-validated models. The parameters used in this figure are $a = 0.98$, $b = 0.46$, $c = 60.50$, $\sigma_\textrm{x} = 2.93$, $\sigma_\textrm{y} = 0.20$, $k = 1.07\times10^{-1}$, $\mu_\textrm{add} = 1.01\times10^{-3}$, and $\sigma_\textrm{add} = 3.04\times10^{-3}$.}
    \label{fig: model validation}
\end{figure}

Before peak-to-peak $V_\textrm{pp}$ detection, we removed the DC component of the electromyography (EMG) recordings and then filtered the high-frequency components of them with a fourth-order Butterworth filter with a cut-off frequency of $600\,\unit{\hertz}$. Additionally, recordings with activity of more than $40$\,\textmu\!V (peak-to-peak) within a window of $200\,\unit{\milli\second}$ before the TMS pulse were marked as facilitated and excluded from the analysis \cite{goetz2018_noise_floor_detection, li2022_noise_floor_detection}. The stimulus strength was expressed as a percentage of the TMS device's maximum amplitude (\% MA).

This study used a self-learning matched-filter algorithm to detect and extract $V_\textrm{pp}$ with increased sensitivity \cite{li2022_noise_floor_detection}. We set the sliding filter detection window to $20\,\unit{\milli\second}$. By visually checking the active MEP responses, the detection window runs from $220\,\unit{\milli\second}$ to $260\,\unit{\milli\second}$, decomposes the active responses into the summation of several motor unit action potentials, and finds the peak-to-peak voltage (see the literature for details \cite{li2022_noise_floor_detection}). Moreover, we applied the same filter to measure the background noise by going through the first $40\,\unit{\milli\second}$ of each MEP measurement.

\subsection{Parameter calibration}
Each subject $i$ has a set of $N_{ij}$ stimulus--response pairs $\{(x_{ijk}, V_{\textrm{pp}, ijk}) \mid k \in [1, N_{ij}], k \in \mathbb{Z}^{+}\}$ for each pulse shape $j$. A specific subject shares the same parameters for $v_\textrm{add}$ across different pulse shapes, and the parameters of $v_\textrm{add}$ are firstly estimated by using the likelihood function of GEV distribution based on the measured background noise. Since the $V_\textrm{pp}$ detection algorithm must generate positive amplitude values for MEP measurements, the GEV distribution should be purely positive. We set up a constraint that forces the distribution to equal zero when the random variable is less than zero. According to Equation \ref{equ: conditional prob equation}, the remaining parameters of $\boldsymbol{\theta}$ (by keeping the parameters of GEV constant) for each subject and each pulse shape are estimated through 
\begin{equation}
    \boldsymbol{\theta}_{ij} = \arg\max_{\boldsymbol{\theta}_{ij}} L(\boldsymbol{\theta}_{ij}) = \arg\max_{\boldsymbol{\theta}_{ij}} \prod_k^{N_{ij}} f_{y|x}(y_{ijk} | x_{ijk}, \boldsymbol{\theta}_{ij})
    \label{equ: objective function}
\end{equation}
under the assumption that the inter-pulse correlation with sufficient temporal spacing is negligible \cite{moller2009_hysteresis_IO}. The likelihood maximisation for the objective function of Equation \ref{equ: objective function} is achieved with a global particle-swarm optimisation algorithm, and all codes are implemented using MATLAB (USA, v2022b). All codes are available online (\nameref{supplementray data}).

\begin{figure*}[ht]
    \centering
    \includegraphics[width=\linewidth]{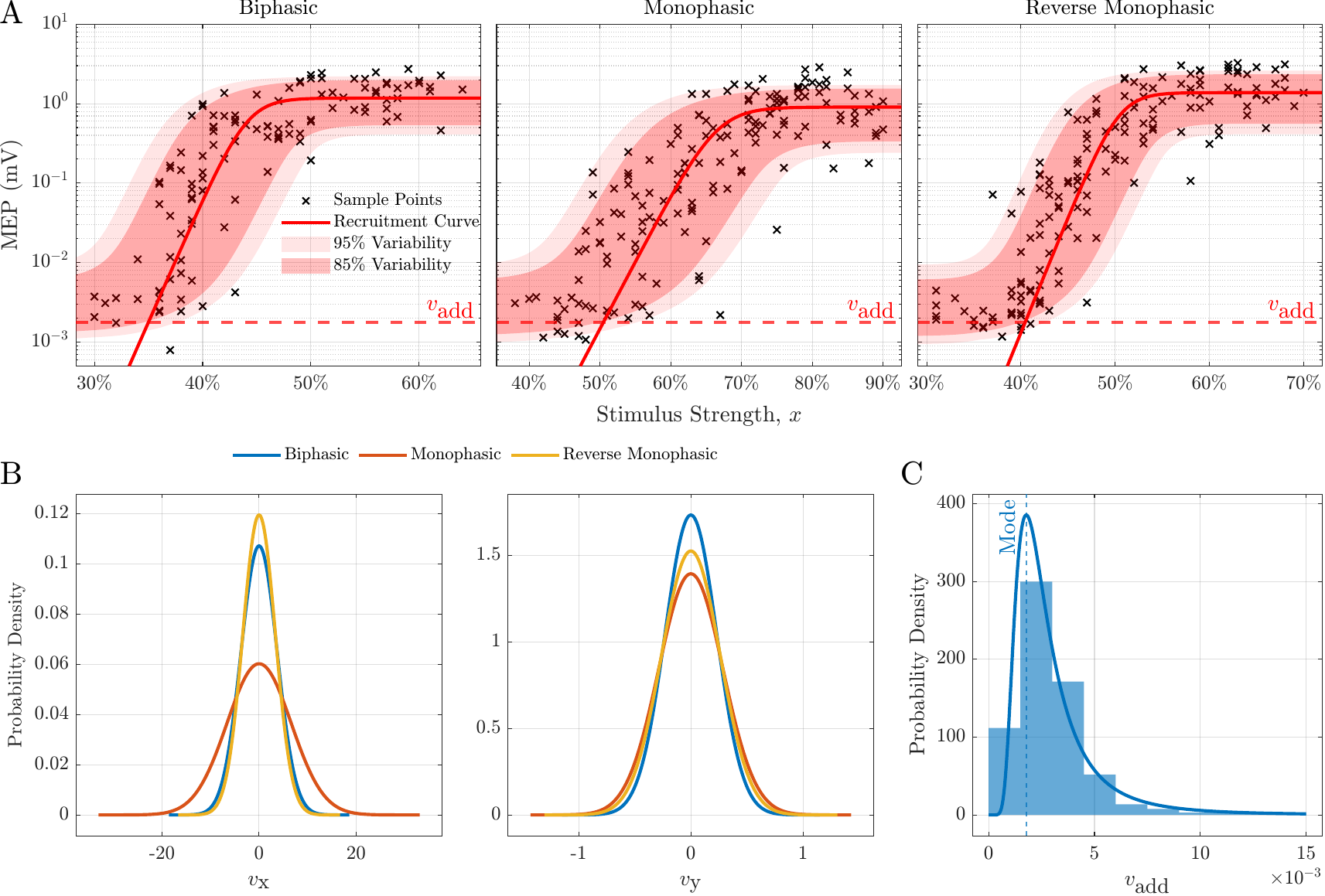}
    \caption{Representative individual IO data, regression results, and distributions of their variability sources for Subject S15A. (\textbf{A}) Illustration of the two components of the triple-variability-source model for each pulse shape, an expected neuronal recruitment curve (red solid line) and a mode value of the background noise $v_\textrm{add}$ (red dashed line); the \textit{x} axis represents the TMS pulse strength (percentage of the maximum amplitude of the device). The \textit{y} axis is the peak-to-peak amplitude of the MEPs. The measured MEP amplitudes to a stimulus are marked with crosses ($\times$). The two variability ranges of Equation \ref{equ: model} are $95\%$ (shaded with light red) and $85\%$ (shaded with moderate red). (\textbf{B}) Distribution of variability sources $v_\textrm{x}$ and $v_\textrm{y}$. (\textbf{C}) Distribution of the GEV model of $v_\textrm{add}$ and the corresponding baseline measurements for the subject; the vertical blue dashed line is the mode value of the GEV model.}
    \label{fig: representative model}
\end{figure*}

\subsection{Model validation, comparison and analysis}
Before calibration on the experimental dataset, we tested the performance of the particle-swarm optimisation algorithm. We simulated the process based on Equation \ref{equ: model} with the theoretical parameters $\boldsymbol{\theta} = [1, 0.45, 60, 3, 0.2, 0.1, 1\times10^{-3}, 3\times10^{-3}]$ and generated $500$ data points with $25$ different stimulus strengths spanning from $30\%$ to $100\%$ of the device's maximum amplitude. There are $20$ measurements for each given stimulus strength $x$. In addition, we used a GEV random number generator to generate $1,000$ background noise data points. To validate the robustness of the optimisation performance and the reliability of model accuracy, we implemented a K-fold cross-validation approach with $K = 30$. In each iteration, 29 subsets were used to calibrate the model. This method assessed the model's generalisability and provided reliable estimates of its performance and parameter stability.

We employed mixed-effect models to analyse the calibrated model parameters, accounting for biological differences among individuals and variations in experimental techniques. For parameters $[a, b, c, \sigma_\textrm{x}, \sigma_\textrm{y}]$, we selected \textit{age} (continuous), \textit{sex} (categorical), and \textit{pulse shape} (categorical) as fixed-effect variables, with \textit{subject} as a random intercept variable. For parameters $[k, \sigma_\textrm{add}, \mu_\textrm{add}]$, we used a different approach. Since pulse shape does not affect background noise, we excluded it from these models. Additionally, the number of subjects equalled the number of sets of GEV parameters, which is insufficient to serve as a random effect. Consequently, we employed simple linear models to analyse these GEV parameters. We used the \textit{lme4} package in R (version 4.3.1) to calibrate the mixed-effect models \cite{lme4}. To test the significance of the fixed-effect variables, we conducted a type-III ANOVA with Satterwaite's method. Moreover, we used the likelihood-ratio test to determine the significance of \textit{subject} for parameters. In total, we developed and analysed eight distinct models: five mixed-effect models for $[a, b, c, \sigma_\textrm{x}, \sigma_\textrm{y}]$ and three linear models for $[k, \sigma_\textrm{add}, \mu_\textrm{add}]$.

Moreover, a previously proposed dual-variability-source model from a decade ago has two variability sources at the input and output sides to fit the MEP IO curve \cite{goetz2012model_MEP_variability}. We calibrated this dual-variability-source model for each subject as well as each pulse shape and used the Bayesian information criterion (BIC) 
as an indicator of goodness-of-fit to support model comparison \cite{wit2012_BIC}. This comparison is appropriate because the BIC takes into account not only the number of parameters of the statistical model but also the quantity of the calibration data. Moreover, we resampled the estimated model parameters and their BIC values using boot-strapping analysis with $1,000$ repetitions for deriving their median and interquartile range. For statistical significance testing, we employed the Mann--Whitney U test for single comparisons and post-hoc analysis. We set significance levels at $\alpha_\text{double} = 0.05$ for double-sided tests and $\alpha_\text{single} = 0.025$ for single-sided tests.

\section{Results}
\subsection{Model validation}

\begin{table}[h]
    \centering
    \caption{Comparison of Theoretical and Estimated Model Parameters}
    \label{tab:param_comparison_horizontal}
    \resizebox{\linewidth}{!}{
        \begin{threeparttable}
            \begin{tabular}{lccccc}
            \hline
             & $a$ & $b$ & $c$ & $\sigma_\textrm{x}$ & $\sigma_\textrm{y}$ \\ \hline
            Theoretical value & 1 & 0.45 & 60 & 3 & 0.2 \\  \hline
            Estimated value & \makecell{0.97 \\ ($1.07\times10^{-2}$)} & \makecell{0.47 \\ ($7.64\times10^{-3}$)} & \makecell{60.32 \\ ($1.48\times10^{-1}$)} & \makecell{2.96 \\ ($4.70\times10^{-2}$)} & \makecell{0.20 \\ ($2.61\times10^{-3}$)} \\
            \hline
            \end{tabular}
            \begin{tablenotes}
            \small
            \item Note: Estimated values are given as median and interquartile range (in the parentheses). The GEV parameters were estimated using simulated baseline background noise and kept the same in each iteration across the cross-validation; they were $k = 1.07\times10^{-1}$, $\mu_\textrm{add} = 1.01\times10^{-3}$, and $\sigma_\textrm{add} = 3.04\times10^{-3}$.
            \end{tablenotes}
        \end{threeparttable}
    }
    \label{tab: model validation}
\end{table}

To clearly present the physiological properties and technical noise of IO curves, we separately plotted the recruitment curves and their corresponding noise floor. Figure \ref{fig: model validation} demonstrates a representative triple-variability-source model calibrated to the simulated dataset. Two variability ranges, i.e., $85\%$ and $95\%$, represent the overall IO trend and encompass most of the simulated data points. The expected neuronal recruitment curve (solid red line) closely follows the trend of the simulated data points and captures the shape of the input--output relationship in responses. The background noise level, represented by the mode value of the variability source $v_\textrm{add}$ (dashed red line), is clearly distinguishable from the recruitment curve. The mode value well represents the background noise because the sample points are denser around the mode value over the entire variable domain.

According to Equation \ref{equ: recruitment curve}, the neuronal recruitment curve almost equals $S(\Tilde{x}) \approx \log_{10}(a)$ when $\Tilde{x}>c$, whereas $S(\Tilde{x}) \approx \log_{10}(a) + {b}\cdot(\Tilde{x}-c)/\ln{10}$ when $\Tilde{x}<c$. Thus, $c$ acts as a junction and divides the neuronal recruitment curve into transition and saturation regions. The curve in the transition region rapidly increases over a small range of stimulus strengths with a slope $b$ from the noise level and becomes saturated at a level of $a$ in the saturation region. Therefore, we can  define $[a, b, c]$ as respectively the saturation level, slope, and shift of the neuronal recruitment curve.

\begin{table*}[t]
    \centering
    \caption{Summary of the estimated IO curve variability and recruitment parameters for the triple-variability-source model}
    \resizebox{\linewidth}{!}{
    \begin{threeparttable}
    \begin{tabular}{l|cccc}
        \hline
         & Biphasic & Monophasic & Reverse Monophasic & ANOVA \\ \hline
         Recruitment curve \\
         \text{   } Saturation level, $a$ & $1.50 \, (1.53)$  &  $0.94 \, (1.22)$    & $1.33 \, (1.15)$ & $p = 0.04$, $F(2, 35.05) = 3.53$ \\
         \text{   } Slope, $b$ & $0.67 \, (0.55)$    &  $0.44 \, (0.30)$    & $0.57 \, (0.25)$ & $p = 2\times10^{-3}$, $F(2,35.56) = 7.41$ \\
         \text{   } Shift, $c$ & $57.84 \, (13.39)$ &  $83.62 \, (14.93)$   & $64.03 \, (14.28)$ & $p = 1.87\times10^{-15}$, $F(2, 35.13) = 103.52$ \\ \hline
         x-variability source, $v_\textrm{x}$ \\
         \text{   } Spread $\sigma_\textrm{x}$ (\% MA) & $3.61 \, (2.12)$   &  $5.98\,(1.81)$  & $4.76\,(2.16)$ &  $p = 1\times10^{-3}$, $F(2,34.77) = 8.43$ \\ \hline
         y-variability source, $v_\textrm{y}$ \\
         \text{   } Spread $\sigma_\textrm{y}$ & $0.17\,(0.17)$ &  $0.27\,(0.20)$  & $0.20 \, (0.13)$ & $p = 0.30$, $F(2,34.80) = 1.23$\\ 
         \text{   } Multiplicative impact on $V_\textrm{pp}$ & $\times/\!\divisionsymbol \,1.65$ & $\times/\!\divisionsymbol \,1.91$ & $\times/\!\divisionsymbol \,1.55$ \\ \hline
         Additive variability source, $v_\textrm{add}$ \\
         \text{   } $k$ & \multicolumn{3}{c}{$0.36\,(0.17)$} & N/A \\
         \text{   } $\sigma_\textrm{add}$ & \multicolumn{3}{c}{$1.1\times10^{-3} \, (4.7\times10^{-4})$} & N/A\\
         \text{   } $\mu_\textrm{add}$ & \multicolumn{3}{c}{$2.3\times10^{-3} \, (7.6\times10^{-4})$} & N/A \\ 
         \hline
         Normalised parameters \\
         \text{   } Slope, $\hat{b}$ & $35.82 \, (24.57)$    &  $31.95 \, (29.95)$    & $33.57 \, (9.89)$ & $p = 0.15$, $F(2, 35.56) = 1.97$  \\
         \text{   } Spread $\hat{\sigma}_\textrm{x}$ (\% MA) & $0.06 \, (0.03)$   &  $0.07\,(0.02)$  & $0.07\,(0.02)$ &  $p = 0.26$, $F(2, 35.02) = 1.40$ \\
         \text{   } Spread $\hat{\sigma}_\textrm{y}$ & $0.23\,(0.17)$ &  $0.22\,(0.16)$  & $0.24 \, (0.16)$ & $p = 0.61$, $F(2,35.15) = 0.51$ \\
         \hline
    \end{tabular}
    \begin{tablenotes}[para] 
        \textbf{Note:} These values are given as median and interquartile range (in the parentheses), both of which are estimated by boot-strapping with $1,000$ repetitions. For the y-variability source, the multiplicative impact on the MEP amplitude, $10^{v_\textrm{y}}$, is reported. For the x-variability source, the spread is given relative to the maximum pulse amplitude of the device (MA). For the additive variability source, each subject has the same set of parameters across different pulse shapes. The normalisation of the recruitment curve is according to the maximum response amplitude along the y-axis and the midpoint of the strength along the x-axis. We used a type-III ANOVA with Satterwaite's method to test the significance of the fixed-effect variable, \textit{pulse shape}.
    \end{tablenotes}
    \end{threeparttable}
    }
    \label{tab: summary of optimised parameters}
\end{table*}

Table \ref{tab: model validation} lists the estimated parameters with their median and interquartile values. The estimated parameters closely match their theoretical value with an average relative deviation on the order of $10^{-2}$ or lower and have narrow interquartile ranges as low as $10^{-3}$, which indicates high accuracy and precision in the estimates. Figure \ref{fig: model validation} (\textbf{B}) contains a probability--probability (P--P) plot with the estimated and theoretical parameters to calculate the corresponding cumulative probabilities of the triple-variability-source model and illustrates that most points align with the diagonal dashed line ($R = 0.995$), which suggests an excellent agreement between the theoretical and the estimated model. In addition, the distribution of the correlation coefficient ($R$) of the P--P plots for these cross-validated models spans a very narrow range from $0.993$ to $0.997$ (inset of Figure \ref{fig: model validation} (\textbf{B})). These results suggest that the optimisation procedure successfully recovered the underlying parameters used to generate the simulated data and validates the optimisation robustness, intrinsic model reliability, and reproducibility with respect to deterministic curve parameters and variability descriptors.

\subsection{Model regression results}
\label{secB: model results}
Figure \ref{fig: representative model} graphs a calibrated model for Subject S15A as a representative example; the data for all subjects is available in the \nameref{supplementray data}. Figure \ref{fig: representative model} (\textbf{A}) demonstrates that the neuronal recruitment curve describes the overall trend of the strength-dependent responses, and its two variability ranges encompass the majority of the sample points. Figure \ref{fig: representative model} (\textbf{B}) shows the distributions of $v_\textrm{x}$ and $v_\textrm{y}$ for different pulse shapes, and Figure \ref{fig: representative model} (\textbf{C}) illustrates that the distribution of the background noise, $v_\textrm{add}$, which forms the low-side plateau, is highly right-skewed and has a long tail towards larger response values. The calibrated GEV model closely matches the distribution of the measurements. In addition, it is an exclusively positive distribution and vanishes below zero. Thus, noise never reduces the MEP amplitude reading in a peak-to-peak detector. This behaviour reflects practical observations.

Moreover, correlation analysis demonstrated a moderate negative correlation between parameters $b$ and $c$ ($R = -0.43$, $p = 8.94\times10^{-4}$), as they are associated and may partly compensate each other in Equation \ref{equ: recruitment curve}. This correlation suggested that an increased value of the motor threshold may decrease the value of the ascending slope. In addition, these recruitment curve parameters are also related to variability sources. The strongest (negative) correlation was observed between parameter $b$ and variability $\sigma_\textrm{x}$ ($R = -0.77$, $p = 2.67\times10^{-12}$). In contrast, parameter $c$ showed a strong positive correlation with $\sigma_\textrm{x}$ ($R = 0.60$, $p = 8.78\times10^{-7}$). These opposite relationships are further supported by the correlation between $b$ and $c$. Additionally, the saturation level $a$ demonstrated a strong negative correlation with variability $\sigma_\textrm{y}$ ($R = -0.73$, $p = 1.56\times10^{-10}$), indicating that higher saturation levels are associated with more consistent muscle unit responses.

Previous research suggested that the curve slope may be a by-product of different thresholds rather than a viable biomarker \cite{peterchev2013_pulse_width_IO}, which aligns with our observed correlation between slope $b$ and shift $c$. The correlation between $b$ and $\sigma_\textrm{x}$ appears to be a consequence of this relationship, suggesting that both slope $b$ and variability $v_\textrm{x}$ are secondary effects of parameter $c$. Likewise, variability $v_\textrm{y}$ would be a secondary effect of parameter $a$. To account for these dependencies and avoid potential misinterpretation, we normalised both $b$ and $\sigma_\textrm{x}$ by $c$ and $\sigma_\textrm{y}$ by $a$ before conducting significance analysis.

Mixed-effects analysis demonstrates that the factor \textit{pulse shape} significantly influences model parameters, while neither \textit{age} nor \textit{sex} show significant impact on any parameter. Table \ref{tab: summary of optimised parameters} summarises the variability sources and recruitment parameters for different pulse shapes derived with boot-strapping. The MEP saturation level varies moderately between pulse shapes ($p = 0.04$, $F(2, 35.05) = 3.53$). Biphasic pulses generate the largest saturation level, comparable to reverse monophasic pulses ($U = 168$, $p = 0.94$). Monophasic pulses with normal current direction demonstrate smaller saturation levels compared to reverse monophasic pulses ($U_\textrm{left} = 118$, $p = 0.03$). Whereas normalisation eliminates differences in the slope $\hat{b}$ ($p = 0.15$, $F(2, 35.56) = 1.97$), the pulse shape significantly shifts the IO curves ($p = 1.87\times10^{-15}$, $F(2, 35.13) = 103.52$). Post-hoc analysis confirms that monophasic pulses with normal current direction shift the curve rightward compared to both biphasic ($U_\textrm{right} = 516$, $p = 1.3\times10^{-6}$) and reverse monophasic pulses ($U_\textrm{right} = 517$, $p = 1.0\times10^{-5}$), reflecting known threshold differences between pulse shapes.

\begin{figure*}[h]
    \centering
    \includegraphics[width = \linewidth]{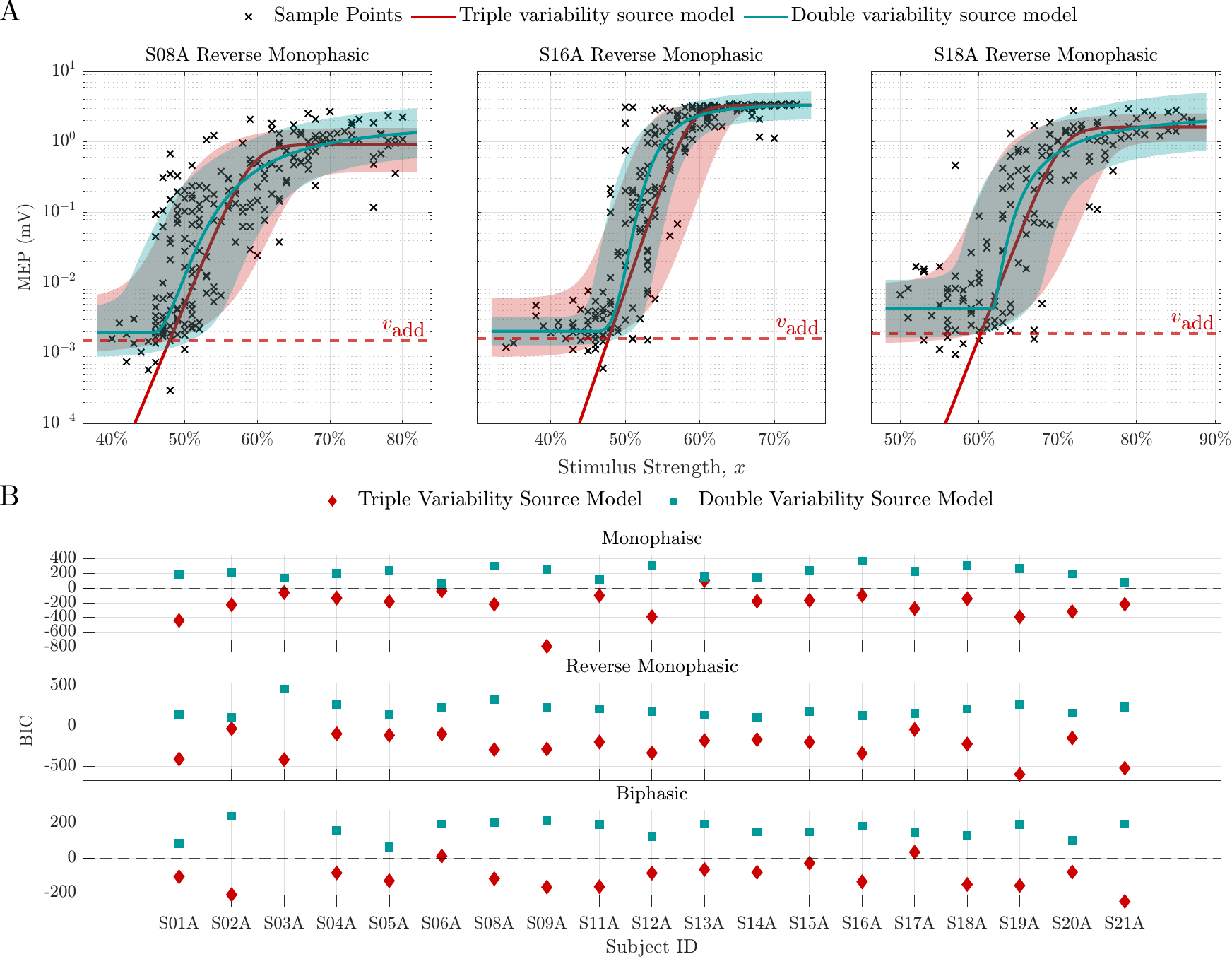}
    \caption{Representative individual IO data and regression results for dual- and triple-variability-source models. (\textbf{A}) Regression results from three selected subjects for both models (red: triple-variability-source model; teal: dual-variability-source model). The x-axis represents the TMS pulse strength as a percentage of the maximum amplitude of the device. The y-axis is the peak-to-peak amplitude of the MEPs. The measured MEP amplitudes in response to stimuli are marked with crosses ($\times$). Each neuronal recruitment curve is associated with a $85\%$ variability range shaded with red (triple-variability-source) or teal (dual-variability-source) colour. (\textbf{B}) Bayesian information criterion (BIC) for both models separately (red diamond: triple-variability-source model; teal square: dual-variability-source model).}
    \label{fig: model comparison}
\end{figure*}

The x-variability source, expressed as a percentage of the maximum TMS device output strength, shows no differences between pulse shapes after normalisation ($p = 0.26$, $F(2, 35.02) = 1.40$). Similarly, the y-variability shows no significant differences for either original ($p = 0.30$, $F(2,34.80) = 1.23$) or normalised values ($p = 0.61$, $F(2,35.15) = 0.51$). The additive variability source, $v_\textrm{add}$, exhibits a right-skewed distribution ($k > 0$) across all subjects, with values concentrated on the left side. With $\mu_\textrm{add} = 2.3\times10^{-3}$ and $\sigma_\textrm{add} = 1.1\times10^{-3}$, the background noise primarily occurs at low magnitudes of $10^{-3}\,\unit{\milli\volt}$ within a narrow range, potentially obscuring MEPs with responses at or below this noise level.

Finally, likelihood-ratio tests reveal significant subject-specific effects on parameters $a$ ($\chi^2 = 23.47$, $p = 1.27\times10^{-6}$), $c$ ($\chi^2 = 26.85$, $p = 2.19\times10^{-7}$), and $\hat{\sigma}_\textrm{y}$ ($\chi^2 = 8.86$, $p = 2.92\times10^{-3}$). These results suggest that saturation level, curve shift, and variability $v_\textrm{y}$ are likely individual features that explain inter-subject variability.

\subsection{Comparison with the dual-variability-source model}

Figure \ref{fig: model comparison} (\textbf{A}) presents three representative regression results for both the dual-variability-source and the triple-variability-source model of three subjects for a visual comparison between them. First, the dual-variability-source model consistently estimates the saturation levels rather poorly compared to the triple-variability-source model. For instance, in the case of the reverse monophasic pulse of S08A (Figure \ref{fig: model comparison} (\textbf{A}), left panel), the tripe-variability-source model saturates for every stimulus strength over $x \approx 60\,\%$, whereas the dual-variability-source curve still increases after $x \approx 80\,\%$. The dual-variability model may also be struggling with poor convergence considering its higher number of degrees of freedom in the sigmoid description.

Moreover, the dual-variability-source model inherently integrates both y-variability sources, specifically the MEP fluctuations and the additive recording noise along the \textit{y} axis, in the same $\sigma_\textrm{y}$, although they differ substantially in size and shape. Thus, the regression results even depend on sampling. For example, regarding the case of S16A (Figure \ref{fig: model comparison} (\textbf{A}), middle panel), the dual-variability-source model estimates a narrow variability range for the noise level because there are more concentrated measurements gathering at the saturation level. Likewise, a higher number of sparse measurements at the noise level likely causes the dual-variability-source model to overestimate the variability range of the saturation level in the case of S18A (Figure \ref{fig: model comparison} (\textbf{A}), right panel). Since background noise and variability as two different effects are lumped together, the sampling, i.e., the number of pulses close to the noise floor vs.\ of pulses at high stimulation strength determines the estimated variability. In contrast, the triple-variability-source model introduces an additional source $v_\textrm{add}$ that specifically accounts for device and physiological noise. Thus, the new model can independently estimate the variability ranges of the saturation and noise levels according to the measurements and precisely capture the distribution differences, as demonstrated in all cases in Figure \ref{fig: model comparison} (\textbf{A}).

Another notable difference is observed in the varying rise slope of the dual-variability-source model, whereas the triple-variability-source model increases with a less steep constant rate in logarithmic scale from the background noise floor before reaching the saturation level. However, in some cases, e.g., S18A (Figure \ref{fig: model comparison} (\textbf{A}), right panel), the motor system responds to the TMS stimulation before the rise of the dual-variability-source model, a scenario well captured by the triple-variability-source model. Additionally, regarding the estimation of background noise, the triple-variability-source model consistently yields lower levels than the dual-variability-source model, with more sample points clustering around the mode value of $v_\textrm{add}$ rather than the low-side plateau in the dual-variability-source model.

Table \ref{tab: comparison of two models} summarises only the variability sources across all pulse shapes and provides the BIC scores for both the dual-variability-source and triple-variability-source models for comparing the goodness-of-fit. The dual-variability-source model estimates variability values similar to those of the triple-variability-source model. However, the regression of IO data with the triple-variability-source model achieves a significantly lower BIC score than the dual-variability-source model ($U = 1,602$, $p = 1.0\times10^{-19}$), which demonstrates a higher descriptive quality without over-fitting. This result suggests that the logarithmic logistic recruitment curve outperforms the original sigmoidal function and better describes the trend of MEP IO curves.

\begin{table}[t]
    \centering
    \caption{Comparison of the estimated IO curve variability and goodness-of-fit for the dual-variability-source model and the triple-variability-source model}
    \resizebox{\linewidth}{!}{
    \begin{threeparttable}
    \begin{tabular}{l|cc}
        \hline
          & Dual variability source model & Triple variability source model \\ \hline
         x-variability source, $v_\textrm{x}$ \\
         \text{   } Spread $\sigma_\textrm{x}$ (\% MA) & $4.46\,(2.56)$ & $4.98\,(2.72)$   \\ \hline
         y-variability source, $v_\textrm{y}$ \\
         \text{   } Spread $\sigma_\textrm{y}$ & $0.22\,(0.12)$ & $0.22\,(0.16)$   \\ 
         \text{   } Multiplicative impact on $V_\textrm{pp}$ & $\times/\divisionsymbol \,1.66$ & $\times/\divisionsymbol \,1.66$ \\ \hline
         Additive variability source, $v_\textrm{add}$ \\
         \text{   } $k$ &  N/A     & $0.36\,(0.17)$   \\
         \text{   } $\sigma_\textrm{add}$ &      N/A    & $1.1\times10^{-3} \, (4.7\times10^{-4})$   \\
         \text{   } $\mu_\textrm{add}$ &     N/A      & $2.3\times10^{-3} \, (7.6\times10^{-4})$  \\ \hline
         Bayesian information criterion & $188 \, (91)$ & $-162 \, (163)$ \\
         \hline
    \end{tabular}
    \begin{tablenotes}[para] 
        \textbf{Note:} These values are given as median and interquartile range (in the parentheses) regardless of pulse shapes, both of which are estimated by boot-strapping with $1,000$ repetitions. For the y-variability source, the multiplicative impact on the MEP amplitude, $10^{v_\textrm{y}}$\!{}, is reported. For the x-variability source, the spread is given relative to the maximum pulse amplitude of the device (MA). For the additive variability source, each subject has the same set of parameters across different pulse shapes.
    \end{tablenotes}
    \end{threeparttable}
    }
    \label{tab: comparison of two models}
\end{table}

\section{Discussion}
Accurate statistical models that can be trained to experimental MEP data and represent the variability accurately can help explain the physiology, extract individual motor system properties, and separate the variability into its components. The variability is not necessarily noise or random. Endogenous signals received by the neuronal target also change the excitability for a short duration and can shift a stimulus--response pair in the IO graph.

\subsection{Comparison with the dual-variability-source model}
Compared with the previous dual-variability-source model, this study used a different recruitment curve without a lower plateau. The lower plateau of the previously typically sigmoidal function was found to not represent a physiological property but rather a technical one, specifically the recording noise floor. Accordingly, this technical property, which would change for an individual based on equipment and set-up, previously led to a different set of IO curve parameters. We therefore separated the low-side plateau and introduced an additional variability as an additive source to represent the background noise. This model could describe varying MEP distributions across the range of stimulus strengths and demonstrate significant improvements over all previous models, including those with already two variability sources.

\subsection{Physiological interpretation of variability sources}
The various variability sources are separated by their different properties and different interactions with the stimulation strength. The different dependence on the stimulation strength indicates different mechanisms and potential locations in the motor system. The variability acting on the input side of the recruitment model with an exclusive effect in the stimulus-strength direction (i.e., along the \textit{x} axis) represents fluctuations of the effective stimulation strength. Such interaction can only occur at the site where the electric field of the TMS coil activates neurons, which then individually respond with an all or none signal, or at places where the signals of various motor units may interact and recruitment information is still available, e.g., the spine, where several motor units can interact synaptically.

Random noise at the membrane of stimulated neurons  \cite{faisal2008_noise_in_neuron, white2000_channel_noise} would be subsumed in this variability term just as any coil position and orientation fluctuations, which would lead to changing electric fields at the stimulation target from stimulus to stimulus \cite{dostilio2016effect, reijonen2020_Coil_MEP, toschi2008reconstruction, koehler2024coil,koehler2023quantitative,goetz2019accuracy}. 
However, these effects are known to not explain the observed variability alone: neither more consistency of coil positioning nor invasive electrical stimulation with spatially fixed electrodes could eliminate the pulse-to-pulse variability of MEP amplitudes \cite{jung2010_navigated_MEP, goetz2022_DBS_MEP_variability, OUPneuronavigation}. This persistent endogenous variability indicates that the \textit{x} variability of MEPs might be dominated by neurophysiological factors such as undulant endogenous excitability changes in the motor system.

The IO data shows that the multiplicative \textit{y} variability cannot be influenced by increasing the stimulus strength and still causes variance along the entire IO curve, including the high-side plateau (Figure \ref{fig: representative model}). This result suggests that this variability is independent of the stimulation strength or recruitment and would not be governed by the cortex excitability \cite{kiers1993_MEP_variability}. Thus, the multiplicative \textit{y} variability source might be attributed to fluctuations in the motor pathway, e.g., from the spinal cord to and in the muscle cells, where the individual units are supposed to have little interaction in healthy individuals \cite{capaday2021_variability_muscle}. Moreover, this particular stimulation-strength-independent \textit{y} is independent of the recruitment, highly skewed but widely normalises under logarithmic transformation, and affects the MEPs all the way to the upper saturation level. Thus, a multiplicative variability term with a simple Gaussian distribution represents this effect well and may indicate that it does not add to but rather modulates an existing signal of an individual motor unit after recruitment in a neuron, a synaptic transmission, or muscle cell. In addition, it is likely subject-dependent and varies among individuals according to the results of likelihood-ratio tests (see \ref{secB: model results}).  The \textit{y} variability should further also include latency fluctuations of signals in the individual motor units from the spinal cord onwards, where they do not interact with each other anymore.

\subsection{Advantages of the triple-variability-source model}
In contrast to the dual-variability-source model, the triple-variability-source model introduces a logarithmic logistic function as the recruitment curve and incorporates an additional, additive variability source after the exponential transformation at the model output (Figure \ref{fig: model structure}). This logarithmic function does not have a lower plateau but continues to fall, which reflects the observation of recent studies that MEP responses appear well below this level but are contaminated and over-shadowed by background noise \cite{li2022_noise_floor_detection,goetz2018_noise_floor_detection}. As in many surface electromyography measurements, the background noise is inevitable and adds independently to the output variability \cite{clancy2002_sampling_noise, boyer2023reducing}. This noise dominates for small MEPs and masks responses below a certain level \cite{li2022_noise_floor_detection,goetz2018_noise_floor_detection}. Benefiting from the absence of a lower plateau, this new recruitment curve provides a mathematical basis for incorporating an additional independent noise floor, allowing it to capture distribution differences between the higher and lower sides (for example, S16A shown in Figure \ref{fig: model comparison}). Table \ref{tab: comparison of two models} demonstrates a better performance of the logarithmic logistic function compared to the original sigmoidal function.

Moreover, this noise is accordingly not a physiological aspect but was accounted for in the multiplicative \textit{y} variability in previous models. The dual-variability-source model describes both the background noise and the fluctuations in the motor pathways, which therefore is a mix of physiology as well as technology and furthermore depends on relatively similar sampling in the low-side and the high-side plateaus \cite{goetz2014_statistical_MEP_model}. The resulting variance is a mix of both and does neither correctly represent the noise nor the physiological \textit{y} variability. Furthermore, the background noise is not only additive but also has a different, very characteristic distribution due to the peak-to-peak MEP detection. Thus, the triple-variability-source model allows for a more accurate and independent estimation of the multiplicative variability inherent in the motor systems as well as the background noise from the perspective of principles of mathematics.

As a result of the combination of various variability effects that share little similarity in one and by wrongly assuming a physiologic low-side plateau, although the IO curve was found to continue falling below that level, the dual-variability-source model lacks robustness when calibrated to data. When the MEP measurements are sparse at the lower stimulus strengths, the dual-variability-source model can entail convergence problems and run into local minima \cite{ma2024_app_IO}. The triple-variability-source model can instead exploit the statistical and causal independence of the background noise for a statistically more stable sequential approach to estimate it and its entire distribution (i.e., the GEV with all parameters) beforehand. Subsequently, it calibrates all remaining model parameters. We demonstrated this approach here.

In addition to the higher robustness, the background noise estimation can use readily available additional data: measuring background noise is relatively easy compared to MEP response measurement. It does not even need TMS pulses administered and can be extracted from as little as several seconds of background recordings before or between stimulation. Alternatively, the triple-variability-source model can also estimate the entire parameter space at once if the MEP IO database is well-recorded and has enough data points along the entire IO curve, i.e., from just background noise all the way into the saturation level.

\subsection{Implications for IO curve biomarkers}
Many studies on MEPs did not systematically consider the entire range of stimulus strengths together but instead calculated their statistics for individual stimulus strengths and performed pairwise statistical tests or similar group analyses \cite{sommer2006_PulseShape_IO,kammer2001motor,goldsworthy2016_TBS_IO}. However, conventional methods may result in spurious results and wrong data interpretation due to unequal and skewed variabilities  \cite{kiers1993_MEP_variability}. The slope, previously reported as a potential biomarker for various aspects, may be an example. A normalisation of the recruitment curve to the parameter $c$, which is similar to converting stimulation strength relative to the motor threshold, eliminated all apparent slope differences (see Table \ref{tab: summary of optimised parameters}). The risk that the absolute slope value may be misinterpreted as an independent biomarker but in many cases might rather reflect threshold differences was observed and reported before in several studies \cite{peterchev2013_pulse_width_IO, goetz2014_statistical_MEP_model}. Normalisation by either the motor threshold or a comparable point on the IO curve eliminated all detectable differences in the slope. Accordingly, the slope may to a large share be a secondary impact of other parameters, i.e., likely a threshold effect, and may therefore not be the best and most robust biomarker for physiological analysis.

\section{Conclusion}
We proposed a new MEP IO model that separates the rather different trial-to-trial variability phenomena along the motor pathway from the cortex through the spine to the muscle into three widely independent sources and provided a framework to extract the parameters from measurement data for an individual subject. Compared to previous models, the triple-variability-source model could more accurately estimate the MEP IO curve and is less prone to bias or over-fitting. Furthermore, as it reliably extracts quantitative information not only about the expected recruitment but also the variability sources, it offers a tool to study the physiology of MEPs, excitation, and neuromodulation. Furthermore, it can serve as a scientific tool to analyse brain stimulation IO curves in clinical and experimental neuroscience without the previous spurious results and statistical bias.


\section*{Supplementary material}
\label{supplementray data}
Supplementary documents to this article can be found online (\url{https://github.com/BIOMAKE/Triple_Variability_Source_IO_Model}).

\bibliographystyle{IEEEtran}
\bibliography{reference.bib}
\end{document}